# Are explicit novelty claims in research abstracts associated with higher ChatGPT scores?

Kayvan Kousha
Statistical Cybermetrics and Research Evaluation Group, Business School, University of Wolverhampton, UK. https://orcid.org/0000-0003-4827-971X;

Mike Thelwall
School of Information, Journalism and Communication, University of Sheffield, UK. https://orcid.org/0000-0001-6065-205X;

Large Language Models (LLMs) such as ChatGPT can estimate research quality from article titles and abstracts, and decision makers may increasingly use LLM scores to support research assessment. In this context, it is a concern that LLM scores may be influenced by linguistic signals rather than the quality of the research described. This study investigates whether explicit novelty claims in research abstracts are associated with higher ChatGPT scores. Using 99,245 journal articles submitted to the UK Research Excellence Framework (REF) 2021, ChatGPT scores and percentile rank gains relative to REF departmental average scores (as a proxy for article-level scores) were compared between abstracts with and without any of 282 three-word phrases identified as indicating a novelty claim. Abstracts containing novelty claim phrases had higher mean ChatGPT quality scores in all four main REF research groupings (panels). Moreover, the score difference was higher for the originality dimension of research quality than for significance and rigour in 30 of the 34 broadly field-based Units of Assessment (UoAs) used to categorise research in the REF. Thus, explicit novelty claims probably associate with higher quality scores primarily because they associate with higher scores for the originality dimension. Follow-up tests suggested that abstracts containing novelty claims tended to rank more favourably under ChatGPT than under the REF departmental quality proxy. Whilst the findings suggest that ChatGPT may be influenced by explicit novelty claims, they do not conclusively show that novelty phrases caused the higher scores because the scores and phrases may at least partially reflect more novel research.

## Introduction

Assessing the quality of large numbers of research publications is costly and time-consuming, creating interest in whether Large Language Models (LLMs) could support this process. Small-scale quality assessments are also common, for example when departments assess candidates’ publications, and there may be a lack of time or relevant expertise for effective evaluations. For these purposes, LLM scoring of articles based on their titles and abstracts can be more practical than processing full texts, giving similar or even more accurate results (Thelwall, 2025a, 2025b; Thelwall & Ali, 2026). Shorter inputs also reduce API processing costs and abstracts are easier to obtain from databases such as Scopus, Web of Science or OpenAlex, whereas full texts may be inaccessible, in varied PDF formats and subject to publisher restrictions on automated processing and copyright compliance. Even when analysing full texts, article abstracts can be particularly influential on the final score (Kaneko, 2026).

The ability to assess research quality from titles and abstracts for large-scale analyses has been found using ChatGPT models (e.g., Thelwall & Yaghi, 2025) and across several medium-sized LLMs (Thelwall & Mohammadi, 2026), supporting their use for large-scale research assessment. One possible reason why research quality scores can be guessed from titles and abstracts alone is that abstracts contain the main information and claims about a study, whereas additional details in full texts may distract current models when assigning scores, although they can support more detailed feedback (Thelwall & Ali, 2026). There is also evidence that LLMs may not use information consistently throughout long documents, particularly when relevant evidence appears in the middle of the input (Liu et al., 2024).

Despite the advantage of LLMs processing titles and abstracts rather than full texts to score academic papers, they can give a misleading impression of a study's value. For instance, an analysis of eight MDPI journals found that claims in science journals were generally expressed more confidently in abstracts than in discussions. Hence, LLMs relying on abstracts may express findings with too much certainty (Thelwall, 2026e). Using International Conference on Learning Representations (ICLR) papers, another study also found that introductions, results and discussions together were most useful for predicting reviewer novelty scores, while full texts did not improve predictions (Wu et al., 2025). Thus, abstract-based scoring can detect originality claims without being able to verify them. This raises the concern that the wording of abstracts may influence LLM assessments more than expert judgements. If so, abstracts could be reframed to gain higher scores, for example by adding unsupported or exaggerated claims that a study is highly novel, makes a significant contribution, has important implications or uses novel methods. The presence of such claims in an abstract therefore does not prove that they are accurate.

There is also some evidence that LLM assessments may be sensitive to how research is presented in journal article abstracts. A study of over 99,000 Research Excellence Framework (REF) 2021 journal articles found that abstract linguistic complexity was more strongly associated with ChatGPT scores than with departmental expert proxy scores in many fields. Similar findings were obtained using individual article quality scores in health and life sciences. This pattern suggests that ChatGPT may be more favourable to complex and longer abstracts than experts (Kousha & Thelwall, 2026b). Moreover, in a study of over 500 unpublished arXiv preprints, "presentation-only revisions" to abstracts, contribution statements and discussions increased AI-review scores without changing the methods, results or other scientific evidence (Yang et al., 2026).

Papers submitted to the International Conference on Learning Representations have also been used to assess whether changes to manuscript wording affect AI-review scores. In a small-scale study, the abstracts of 80 rejected conference papers were changed. These included 40 human-authored papers from ICLR 2025 and 40 AI-generated papers from Agents4Science 2025. AI-review ratings increased when abstracts were rephrased, framed to emphasise novelty and contribution, or made to overclaim. A further analysis of 20 Nature Communications papers showed that abstract rewriting also increased AI-review ratings (L. Li et al., 2026). Similar findings were reported in a larger study of 640 arXiv manuscripts that had not been accepted to five 2025 conferences (including ICLR 2025). The abstracts (only) were repeatedly paraphrased using feedback from LLM review scores while aiming to preserve their meaning. Ratings from three LLMs increased when the full papers were assessed with the remaining text unchanged (Kaneko, 2026). In another study of 120 ICLR 2026 submissions, full manuscripts were rewritten by LLMs (GPT-5.5 and Claude Opus 4.8) to make novelty claims seem more assertive or more cautious. Five LLM reviewers scored the papers on a 1–10 scale. Under the standard review prompt, cautious novelty framing generally lowered AI review scores relative to the original manuscripts, whereas assertive framing generally increased scores, although the effects varied across the models (M. Li et al., 2026). These findings suggest that LLM review scores can be sensitive to how research is framed in abstracts and full manuscripts. Nevertheless, the effects can vary by model, and most of this evidence comes from conference papers in machine learning and related fields, so it is not safe to extrapolate from these findings to the rest of science.

LLM assessments may also be influenced by broader contextual factors that do not necessarily reflect the quality of an individual study. LLM scores have been associated with publication year, field and first-author country, even when these factors were not explicitly provided to the models for assessment (Thelwall & Kurt, 2025). In a controlled study, ChatGPT-4o mini gave lower quality scores to identical manuscripts when they were attributed to less prestigious institutions (Howell et al., 2025). Possible indirect gender-related differences have also been found in ChatGPT assessments based on titles and abstracts. Although author details were not provided in another study, male first-authored papers received slightly more favourable ChatGPT scores relative to expert proxy scores (Kousha & Thelwall, 2026a). Topic-related differences in LLM scoring have also been found, for example, for virus and machine-learning research, although the topics receiving relatively higher scores varied between LLM models (Thelwall, 2026c). LLMs may also fail to recognise important weaknesses of published research. ChatGPT quality reports did not mention the retraction status or relevant problems of retracted articles (Thelwall et al., 2025) and other LLMs mostly failed to identify high-profile retracted articles from their titles and abstracts (Thelwall, 2026a). Thus, LLM scores may vary with factors beyond the quality of an individual study and may overlook serious problems in the research.

## Originality in research assessment

Originality is one of the three most common dimensions of research quality used in assessments (Langfeldt et al., 2020), alongside significance and rigour. The REF also uses these three dimensions, with originality defined as:

> *"the extent to which the output makes an important and innovative contribution to understanding and knowledge in the field. Research outputs that demonstrate originality may do one or more of the following: produce and interpret new empirical findings or new material; engage with new and/or complex problems; develop innovative research methods, methodologies and analytical techniques; show imaginative and creative scope; provide new arguments and/or new forms of expression, formal innovations, interpretations and/or insights; collect and engage with novel types of data; and/or advance theory or the analysis of doctrine, policy or practice, and new forms of expression."* (Research Excellence Framework [REF], 2019, p. 34)

The REF definition of originality above applies across all four of the main panels that organise research assessment, although their supplementary criteria give different disciplinary examples relevant to originality. Panel A (health and life sciences) describes characteristics that can apply at different star levels, including "significant addition to knowledge and to the conceptual framework of the field". For four-star outputs, Panel B (engineering, physical sciences and maths) refers to "new thinking, new techniques or novel results" and "developing new paradigms or fundamental new concepts for research". Panel C (social sciences) refers to work that is "outstandingly novel in developing concepts, paradigms, techniques or outcomes". Panel D (arts and humanities) includes work that is "outstandingly novel, innovative and/or creative" and "instrumental in developing new thinking, practices, paradigms, policies or audiences" (REF, 2019, pp. 35–40). Hence, the concept of research originality differs across disciplines, which should be considered when evaluating across broad subjects.

New knowledge and innovation also appear in other national research assessment frameworks. For instance, Excellence in Research for Australia (ERA) defines research as "the creation of new knowledge and/or the use of existing knowledge in a new and creative way so as to generate new concepts, methodologies, inventions and understandings" (Australian Research Council, 2019). The Dutch Strategy Evaluation Protocol 2027–2033 also refers to "theoretical innovation, such as proposing new questions and perspectives that can challenge established paradigms" (Universities of the Netherlands et al., 2026, p. 9). These examples indicate that different research assessment exercises may define originality or novelty in slightly different ways, but it is a core concept.

The use of LLMs to assess originality has also been examined. In one study, six LLMs assessed 965 biomedical papers across two datasets by comparing their titles and abstracts with those of five earlier related papers. One dataset compared Nobel Prize papers with other papers by the same laureates as well as random papers. Another used a citation-based disruption index to identify disruptive, developmental and general papers. The models distinguished originality levels to some extent but appeared to be “overly lenient” (Huang et al., 2025). In a study of 139 research proposals submitted to the Austrian Science Fund, ChatGPT-4o was used to compare pairs of proposals and select which should be funded. Proposals rated as more novel by human reviewers were more likely to be funded by humans but not by the LLM comparisons (Machado, 2026).

Although some studies have tested how rewriting conference papers affects LLM review scores, no large-scale study has investigated whether originality or novelty (here used as synonyms) claims already present in published journal article abstracts are associated with higher LLM quality scores, whether this association differs by novelty phrase, across disciplines, or how it compares with expert judgments. The following research questions are addressed to fill these gaps:

1. Do ChatGPT overall-quality and originality scores differ between articles whose abstracts contain explicit novelty claims and those whose abstracts do not?
2. Does the difference between ChatGPT overall-quality scores and REF departmental proxy scores vary between articles with and without such claims?
3. Do these differences vary across REF main panels and Units of Assessment?

# Methods

This study used the titles and abstracts of journal articles submitted to the UK’s most recent national research assessment: REF 2021. ChatGPT overall quality and originality scores were based on the articles’ titles and abstracts. REF2021 departmental average scores were used as proxies for article quality scores in the absence of article-level REF scores. To identify explicit originality claims, a dictionary was developed through LLM-assisted selection of three-word phrases (3-grams) extracted from the abstracts, followed by manual checks by the authors. Articles with and without identified claims were compared using both ChatGPT scores and the REF proxy scores. The novelty claims identified in abstracts were used as evidence about how the research was described, rather than evidence that the research was genuinely original.

### *Dataset*

The source dataset contained 99,245 unique journal articles submitted to REF 2021. The data were obtained from a previous large-scale study of research quality assessment using ChatGPT (Thelwall, 2026d). Sample sizes varied across analyses because score availability differed and articles submitted to more than one panel or Unit of Assessment (UoA) were counted separately within each relevant group. For example, the same article could be included in UoA 1 (Panel A) and UoA 7 (Panel B), or in several UoAs within the same panel.

The composite panel-level overall-quality indicator, derived from different ChatGPT models, was available for 98,820 records, whereas the ChatGPT-5 mini structured-score analyses of originality, significance and rigour used 99,249 article-panel records. At the UoA level, the structured-score analysis used 106,833 article-UoA records, while the rank-gain analysis used 106,051 records with available composite ChatGPT overall-quality and REF proxy scores.

The overall-quality indicator was the mean of the available overall scores from ChatGPT-4o, ChatGPT-4o mini and ChatGPT-5 mini. The originality indicator was the separate originality score from a structured ChatGPT-5 mini REF quality score assessment. The REF proxy score for each article was the mean quality score of the journal article outputs submitted by the relevant department(s). These departmental means had been calculated from individual REF scores before they were deleted (Thelwall et al., 2023). Hence, the REF proxy scores reflect the average quality of a department's submitted journal articles, rather than an expert rating of each article, because these individual scores were deleted by the REF as part of its privacy policy.

**ChatGPT and REF scores**

ChatGPT scores were generated from article titles and abstracts through the OpenAI batch API using zero-shot prompts based on the REF 2021 criteria from the relevant Main Panel (Thelwall, 2026d). ChatGPT-4o and ChatGPT-4o mini each assessed articles five times. ChatGPT-5 mini used five overall-quality assessments and one structured assessment reporting originality, significance, rigour and overall quality. ChatGPT overall quality scores and REF departmental proxy scores were converted separately into percentile ranks. ChatGPT was asked to assess articles using the REF 1*–4* quality criteria, ranging from nationally recognised to world-leading research. Scores from five repeated assessments were averaged across models and so the resulting ChatGPT scores could be fractional. For each article, its ChatGPT rank gain was calculated by subtracting the REF proxy percentile rank from the ChatGPT percentile rank. A positive gain indicates a more favourable ChatGPT score than REF (departmental proxy) score. Rank gains were then compared between articles with and without novelty claims.

**Extraction of novelty claim phrases**

Phrases indicating novelty claims were systematically identified through a multi-step procedure. First, copyright information and structured abstract headings, such as "Originality/value", were removed from the abstracts before processing. These cleaned abstracts had also been used for ChatGPT scoring. This data cleaning prevents a required heading or copyright text not written by the author from being counted as an author's originality claim. A program was then used to extract all sequences of three consecutive words (3-grams) from the abstracts. The script converted text to lower case, replaced hyphens and punctuation with spaces, and kept stop words such as "of", "the" and "not" (e.g., "of the first" and "not previously reported"). It counted each phrase once per abstract, even when occurring more than once. Since this produced a very long list, only the most frequent phrases were kept, with 50 being selected heuristically as the cutoff overall, and 20 as the cutoff for all the words within a Main Panel (to allow field-specific common phrases that are rare in other fields). Combining these five lists and removing duplicates, this identified 30,303 3-grams commonly occurring in REF2021 article abstracts. These two heuristically chosen cutoffs (20 and 50) are a practical limitation because the analysis is based on reasonably common novelty indicating phrases, excluding rarer versions. To check whether LLM screening excluded potentially relevant phrases, the three-word phrases not selected by Gemma 3 were searched for the main novelty terms. Only 43 (0.15%) contained "first", 22 (0.07%) contained "new", 29 (0.10%) contained "previous" or "previously", and none contained "novel", "novelty" or "innovation". In comparison, among the 282 selected phrases, 59 (20.9%) contained "first", 106 (37.6%) contained "new" or "newly", 79 (28%) contained "novel", "novelty", or "innovative", and 38 (13.5%) contained other originality-related wording (e.g., previously, unique, original). This check identified relatively few excluded phrases containing the selected novelty terms, although their use as novelty claims was not assessed in context.

**LLM screening and manual checking**

The 30,303 unique three-word phrases occurring in at least 50 abstracts overall or at least 20 abstracts within a main panel were submitted individually to the Gemma 3 12B LLM, which was chosen for good performance on academic text processing tasks (Thelwall & Mohammadi, 2026) and for being more efficient than larger models for this short simple task. The model was asked whether each phrase was a useful linguistic indicator that a sentence contained an academic novelty claim. Gemma 3 was prompted to assess the evidence provided by the phrase itself and to report a binary decision about whether it was a likely novelty indicator. The full prompt, including examples, is reproduced in Appendix A. These 30,303 unique three-word phrases were screened using Gemma 3 12B, producing 810 phrases that this LLM assessed as being indicative of a novelty claim. Because LLMs are unreliable, the 810 candidate phrases were independently assessed by both authors to identify novelty-suggesting phrases. A score of 1 indicated a clear novelty indicator (e.g., "the first attempt", "study provides new", or "an innovative approach"), 0.5 indicated a possible or ambiguous novelty claim (e.g., "this article contributes", "we experimentally demonstrate" or "this article advances"), and 0 indicated that the phrase was not a useful indicator (e.g., "a recently developed", "the first one", or "was not observed"). Both authors classified 282 phrases (34.8%) as clear novelty indicators and agreed that 444 phrases (54.8%) should not be included because they were rated 0 or 0.5. They disagreed on the remaining 84 phrases (10.4%), with one author rating them for inclusion (1) and the other for non-inclusion (0 or 0.5). Hence, the authors agreed on 89.6% of the classifications with high agreement beyond chance (Cohen's $\kappa = 0.79$, $p < .001$). Only the 282 phrases rated 1 by both authors were included in the final analysis. These phrases were judged by both authors and one LLM to be useful indicators of novelty, although in some cases their interpretation may depend on context. Thus, these phrases were treated as linguistic indicators of how the authors had described their research, rather than as evidence that the research described was genuinely original.

### Matching the phrases to abstracts

The 282 novelty-indicating phrases were automatically matched against normalised abstract text to identify which abstracts contained at least one of the 282 phrases. The phrases and abstracts were converted to lower case, punctuation was replaced with spaces and repeated spaces were removed. Matching was conducted within sentences and required the same consecutive sequence of complete words. Each abstract was counted only once in the main comparison, regardless of how many phrases it contained. Individual phrases were assessed separately in the secondary exploratory analyses. The complete list of 282 accepted phrases is in Supplementary Table S1.

## Results

### Novelty phrases in abstracts and ChatGPT scores

Of the 98,820 abstracts with a ChatGPT overall-quality indicator, 17,257 (17.5%) contained at least one of the 282 identified novelty phrases and 81,563 (82.5%) did not. The ChatGPT score used was the average of the overall quality scores from ChatGPT-4o, ChatGPT-4o mini and ChatGPT-5 mini, since combining LLM scores usually gives stronger results, even if some of the LLMs are weaker. Overall, abstracts containing at least one of the 282 novelty phrases received higher mean overall-quality scores from ChatGPT in all four REF panels. The differences were all statistically significant and ranged from 0.080 points in Panel A to 0.106 in Panel C (Figure 1).

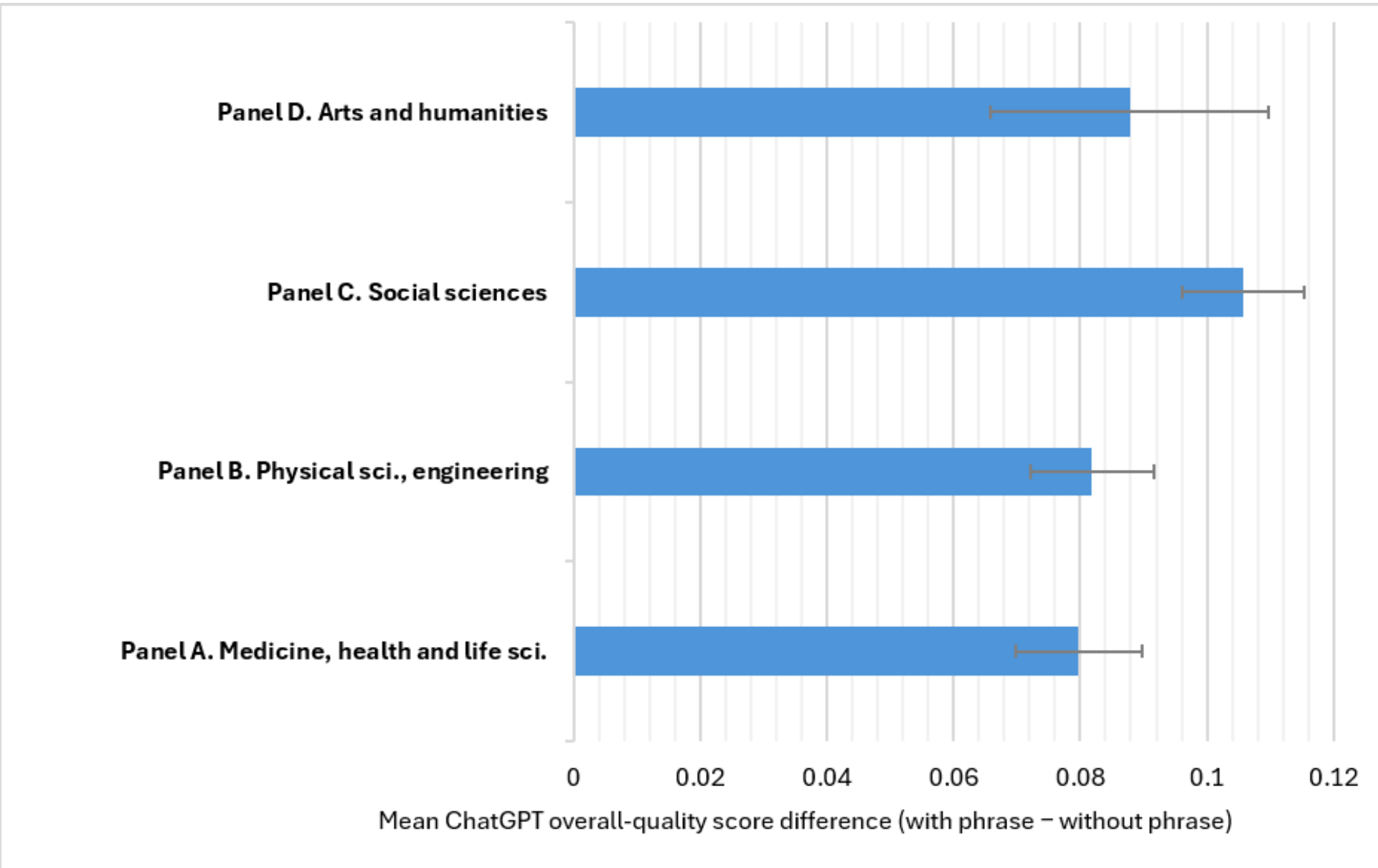

Figure 1. Mean difference in ChatGPT overall-quality scores between abstracts with and without one of the 282 novelty phrases by REF main panel (N=98,820). Error bars show 95% confidence intervals.

The differences in Figure 1 do not show that the novelty phrases caused the higher scores because other characteristics of the abstracts or underlying research may also have contributed. Hence, a further analysis compared the ChatGPT-5 mini structured originality, significance and rigour scores (Figure 2). The originality score difference was larger than the significance and rigour score differences in all four panels. This difference is largest in Panel A, where the differences were 0.116 for originality, 0.023 for significance and 0.002 for rigour, and Panel B, where they were 0.103, 0.027 and −0.009, respectively. Panel C had positive differences for all three criteria, but originality still had the largest difference (0.133). The differences in Panel D were 0.074, 0.053 and 0.037. These results suggest that the score advantage was strongest for originality, particularly in Panels A and B, rather than being an equal increase across all three REF research quality criteria. Thus, explicit novelty phrases had a stronger association with higher ChatGPT originality scores than significance and rigour scores, especially in Panels A and B.

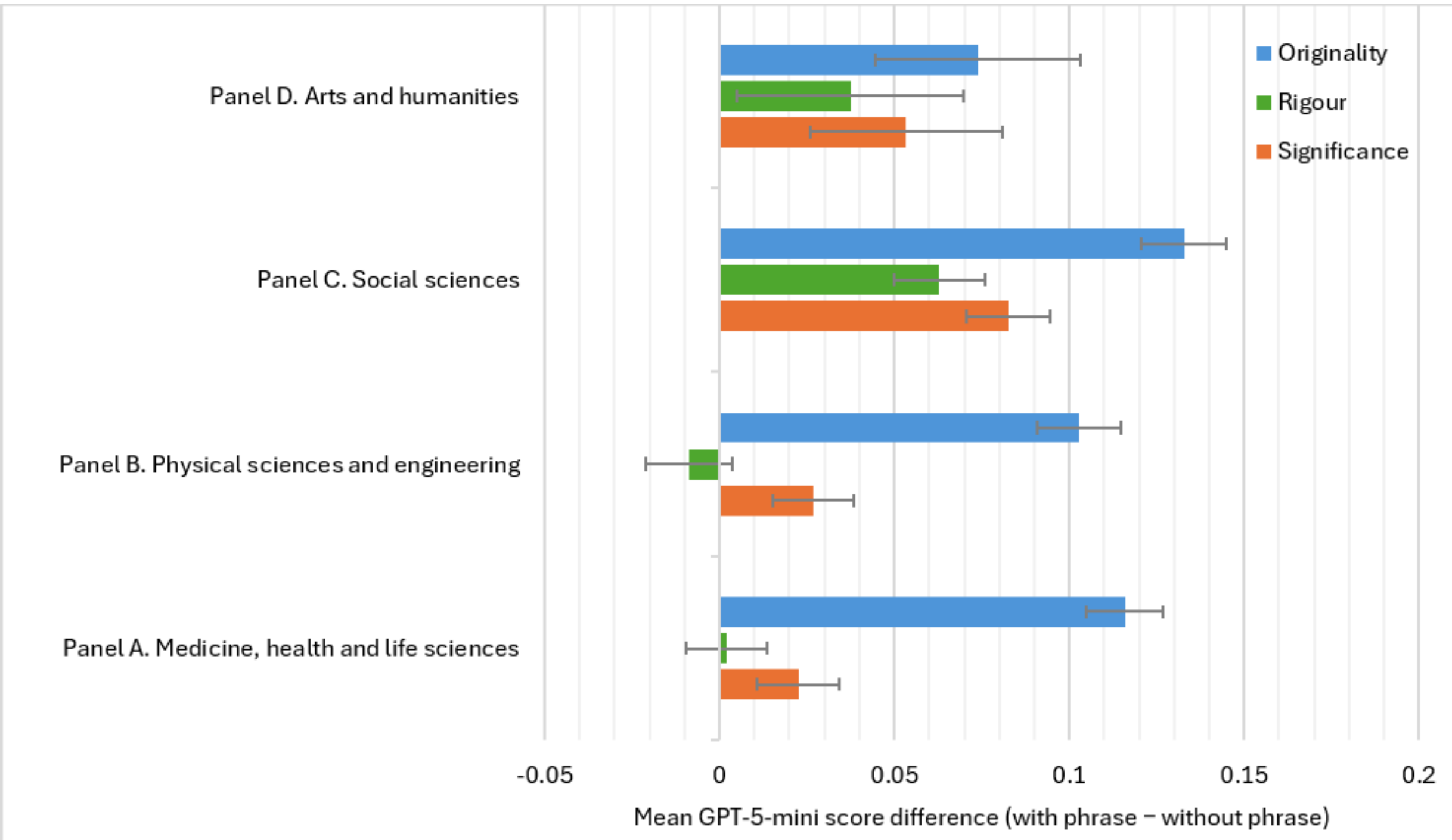

Figure 2. Mean differences in ChatGPT-5 mini originality, significance and rigour scores between abstracts with and without the 282 novelty phrases by REF main panel (N=99,249 article–panel records). Positive values indicate higher scores for abstracts containing the phrases. Error bars show 95% confidence intervals.

**ChatGPT originality score differences by Unit of Assessment**

Across the 34 UoAs within the four Main Panels, the mean ChatGPT-5 mini originality score difference between abstracts with and without accepted novelty phrases was larger than both the corresponding significance and rigour score differences in 30 UoAs (Figure 3). The four exceptions were UoAs in arts and humanities (25, 29, 31 and 33), where the sample sizes are small. The originality error bar also did not overlap with either of the other two error bars for the corresponding significance and rigour scores in 10 UoAs. This includes all six UoAs in Panel A, UoAs 7 and 12 in Panel B, and UoAs 17 and 24 in Panel C. For example, in UoA 3 (Allied Health Professions, Dentistry, Nursing and Pharmacy), the originality difference was 0.186 points, compared with 0.080 for significance and 0.049 for rigour. The corresponding differences were 0.183, 0.082 and 0.037 points in UoA 12 (Engineering), and 0.227, 0.130 and 0.053 points in UoA 24 (Sport and Exercise Sciences, Leisure and Tourism). In UoA 1 (Clinical Medicine), abstracts containing the novelty phrase had a higher originality score by 0.077 points but negative significance and rigour score differences.

Whilst the above results show that ChatGPT-5 mini scores abstracts higher (especially for originality) when they include a common novelty claim phrase, they do not show that the wording caused the higher scores, since the novelty claims (and lack of novelty claims) may be accurate rather than linguistic choices for framing purposes. Papers containing these phrases may have reported genuinely original research, such as new methods, novel findings or original topics. Other differences in research topic, abstract style or research quality may also have contributed (Kousha & Thelwall, 2026b; Thelwall, 2026c). Nevertheless, since ChatGPT assessed only the titles and abstracts and probably could not verify the claims against the full articles (Thelwall, 2026a, 2026b), this does not verify that the AI research quality assessments were completely unbiased. For example, previous studies showed that LLM peer-review scores can increase after changes to abstract wording or presentation without changes to underlying scientific content (L. Li et al., 2026; Yang et al., 2026).

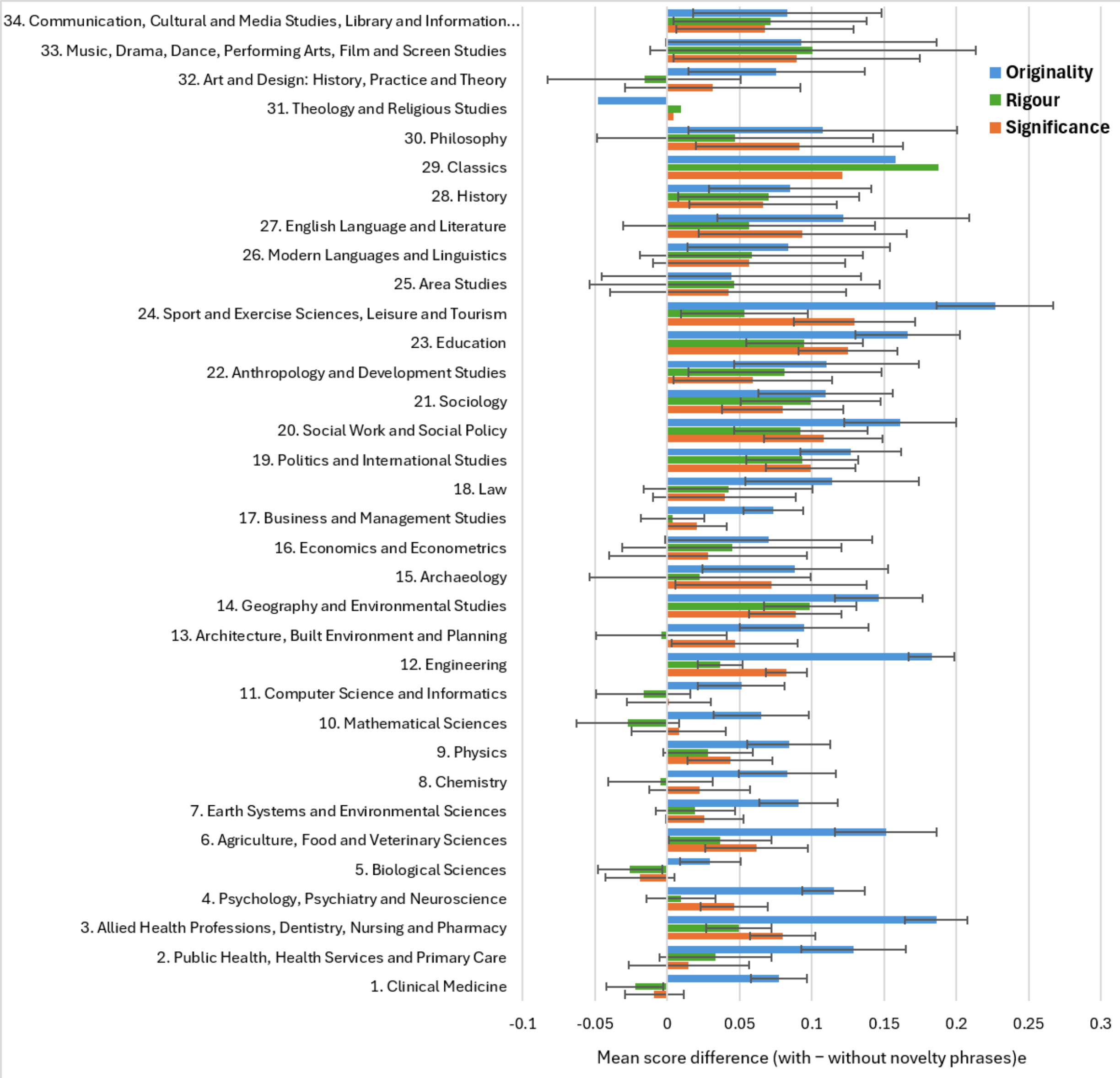

Figure 3. Mean differences in ChatGPT-5 mini originality, significance and rigour scores between abstracts with and without accepted novelty phrases across the 34 REF Units of Assessment (N=106,833 article–UoA records). Positive values indicate higher mean scores for abstracts containing an accepted phrase. Error bars show approximate 95% confidence intervals. Error bars are omitted for UoA 29, Classics (14 abstracts with phrases), and UoA 31, Theology and Religious Studies (11 abstracts with phrases), because their extremely wide intervals would reduce the visual comparability of the other UoAs.

### ChatGPT rank gain relative to the REF proxy

Figure 4 compares ChatGPT rank gains compared to the REF departmental proxy (average department score) between abstracts with and without accepted novelty phrases across the 34 UoAs. ChatGPT overall quality scores and REF departmental scores were converted into percentile ranks within each UoA so that their relative positions could be compared on the same scale. This is because the ChatGPT score was for an individual article, whereas the REF proxy was the related departmental average score for its journal articles submitted to that UoA. ChatGPT rank gain was calculated by subtracting the REF proxy percentile rank from the ChatGPT percentile rank. Thus, a positive value indicated that an article was ranked more favourably by ChatGPT than by the REF proxy. Mann–Whitney U tests were used to compare rank gain between abstracts with and without accepted phrases within each UoA. For example, in Clinical Medicine, the 1,582 abstracts containing an accepted novelty claim phrase had a mean rank of 5,160 compared with a mean rank of 4,731 for the 8,020 abstracts without a novelty claim phrase. This mean rank difference of 429 was statistically significant ($p < .001$), indicating that abstracts with novelty phrases tended to have more favourable ChatGPT rank gains compared with their REF proxy scores.

Abstracts containing the common novelty phrases had higher mean ChatGPT rank gains in 33 out of the 34 UoAs and these differences were statistically significant in 26 UoAs. Six of the eight non-significant UoA differences were from Panel D, where the numbers of abstracts having novelty phrases were often small. For instance, only 10, 14, and 43 of the abstracts in Theology and Religious Studies, Classics, and Area Studies had novelty phrases. In contrast, the differences were statistically significant in all six Panel A UoAs, five of the six Panel B UoAs and 11 of the 12 Panel C UoAs. The only non-significant results in panels B and C were for Mathematical Sciences ($p=0.052$) and Economics and Econometrics ($p=0.059$), respectively. Engineering had the largest raw mean-rank difference, but this should not be interpreted as the strongest association because the magnitude of raw mean ranks is affected by the number of records within each UoA and so the results should be interpreted mainly by direction of mean rank differences and statistical significance.

Although the results suggest that abstracts containing novelty phrases tended to receive more favourable ChatGPT research quality scores relative to the REF departmental proxy, they do not show that ChatGPT disagreed with expert assessments of individual articles because individual REF scores were unavailable. They also do not show that the novelty phrases caused the differences, since research topics, methods, writing style and other features of the titles and abstracts may also have influenced the scores.

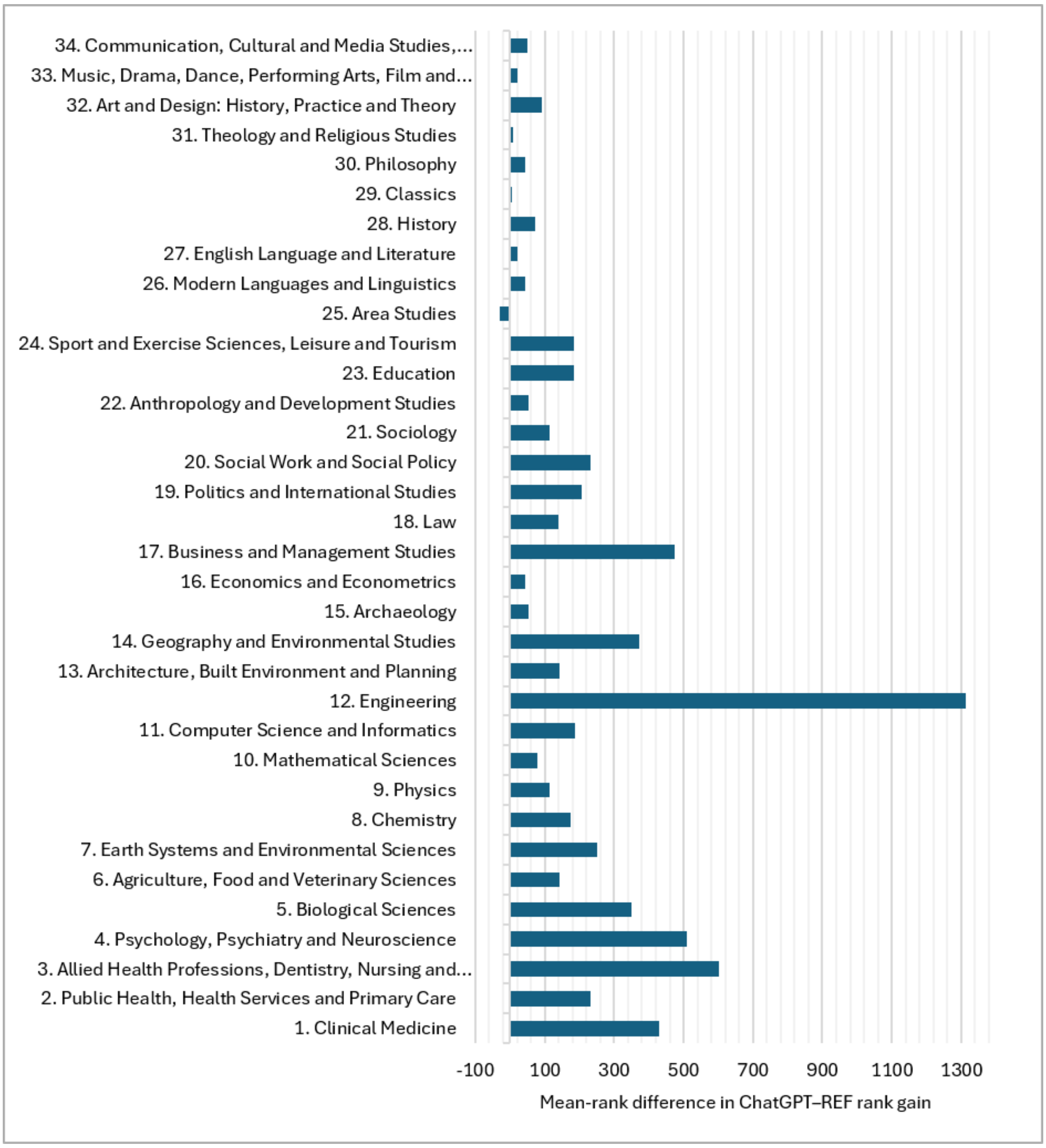

Figure 4. Mann–Whitney mean-rank differences in ChatGPT rank gain relative to the REF departmental proxy between abstracts with and without a novelty phrase across REF UoAs (N=106,051). Positive values indicate greater ChatGPT rank gain for abstracts having a novelty phrase.

## Discussion

The results are limited by covering a single country and year range and by focusing only on journal articles. The biggest limitation is the use of the REF department average quality proxy due to the lack of individual article scores. The results may therefore relate to departmental patterns (e.g., if some departments had abstract writing training sessions that influenced their authors' writing styles) rather than article-level patterns. The 282 novelty phrases used in this study may have missed less frequent claims or forms of novelty expressed differently across disciplines. Some matched phrases may occur without expressing a novelty claim about the authors' own research, since individual occurrences were not checked in context. This may be particularly important in Panel D, where the REF criteria refer to work that is "outstandingly novel, innovative and/or creative" or "instrumental in developing new thinking, practices, paradigms, policies or audiences" (REF, 2019, pp. 35–40), rather than only a new method or first study. Abstract length may also contribute to the observed associations because longer abstracts provide more opportunities to contain an accepted phrase and may receive higher ChatGPT scores (Thelwall & Kurt, 2025; Kousha & Thelwall, 2026b). The study also excluded books and monographs, which are major outputs in many arts and humanities fields. ChatGPT scores were based only on titles and abstracts which may not provide enough evidence to assess the originality of the full research. It is unclear whether using full texts would produce different results. The findings may also depend on the specific ChatGPT models and prompts used.

The results broadly align with the small-scale results discussed above for ICLR conference papers, where manipulated originality claims in abstracts were associated with higher LLM scores (L. Li et al., 2026; M. Li et al., 2026; Kaneko, 2026) and extend them from computer science to almost all academic fields as well as to journal articles.

**Are ChatGPT originality scores higher for abstracts with multiple novelty phrase families?**

A follow-up analysis investigated whether ChatGPT-5 mini originality scores were higher when abstracts contained phrases from more than one novelty family: firstness, novelty/innovation, new contribution or other novelty. Among the 17,347 abstracts containing at least one accepted novelty phrase, 1,381 (8.0%) contained phrases from two or more novelty families. Each family was counted once per abstract, regardless of how many accepted phrases from that family matched. For example, "for the first" and "the first time" counted together as one family (firstness), whereas an abstract also containing "a novel approach" counted as containing two families. Across all four REF panels, abstracts containing phrases from one novelty family had higher mean originality scores than those having no novelty phrase. Scores were higher again for abstracts having phrases from two or more novelty families, and these had the highest mean scores in every panel (Figure 5). Nevertheless, this could again reflect either ChatGPT's language sensitivity or genuinely identifying more original research. For instance, an abstract might claim to be "the first study" to use "a novel method" that produces "a new treatment". In this case the combined presence of these phrases suggests more originality than any single phrase, but the presence of multiple families does not necessarily indicate multiple independent original contributions.

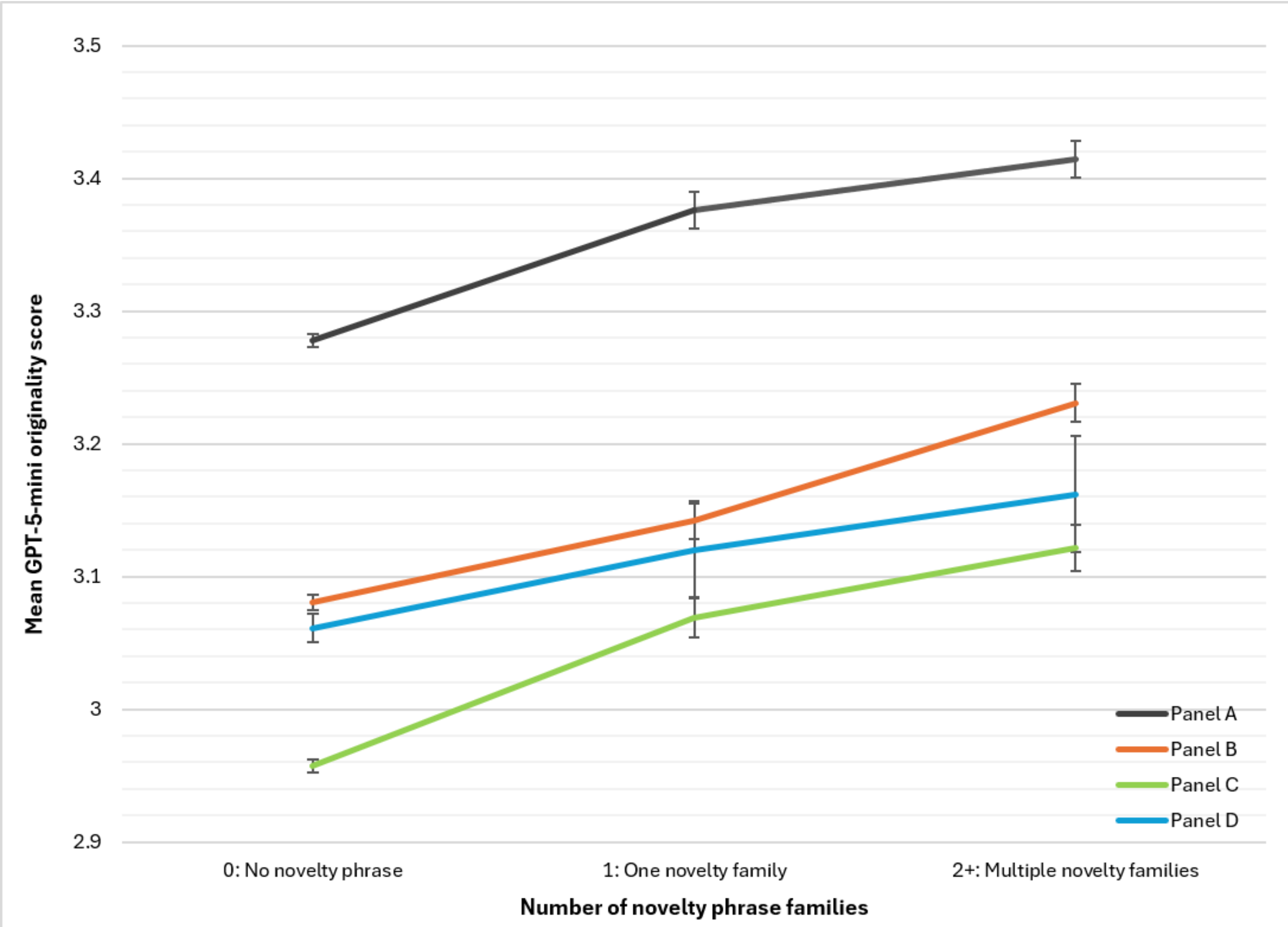

Figure 5. Mean ChatGPT-5 mini originality scores for abstracts containing phrases from zero, one, or two or more novelty families, by REF main panel (N=99,249). Error bars show 95% confidence intervals.

**Phrase-level differences in ChatGPT originality scores**

Among the 17,347 abstracts containing at least one accepted novelty claim phrase in the dataset, 64 phrases occurred at least 100 times. For each phrase, the mean ChatGPT-5 mini originality score was compared with the mean for abstracts containing at least one accepted phrase but not the phrase being assessed. For example, abstracts containing "report the first" had a mean score of 3.577, compared with 3.229 for abstracts containing other accepted novelty phrases but not this exact phrase ("report the first"), giving a difference of +0.348. Figure 6 shows the 20 largest positive differences. Other large differences were found for "present the first" (+0.280), "a novel mechanism" (+0.279), "provide the first" (+0.267) and "the first evidence" (+0.239). In contrast, the two most frequent phrases, "the first time" (N=2,153) and "for the first" (N=2,119) had smaller differences of +0.106 and +0.094, respectively. Score differences varied across phrases. However, this may reflect differences in subject area, research topic, genuine originality or other novelty phrases in the same abstract, rather than the wording itself.

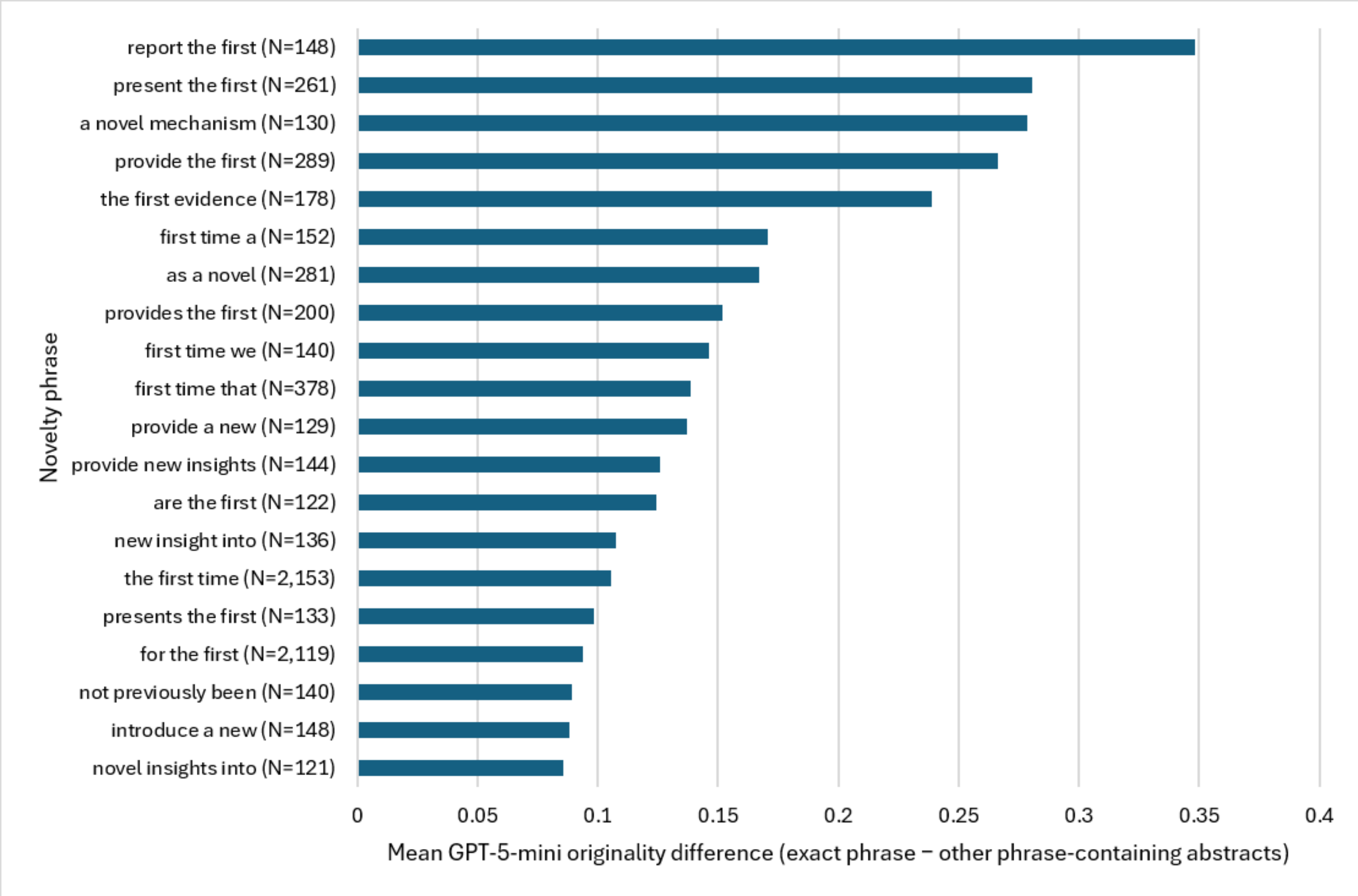

Figure 6. The 20 novelty phrases with the largest positive differences in mean ChatGPT-5 mini originality scores compared with other abstracts containing an accepted phrase but not that exact phrase. Only phrases occurring in at least 100 abstracts were assessed; frequencies are shown in parentheses.

## Conclusions

In answer to the first research question, abstracts having at least one accepted novelty phrase received higher mean ChatGPT overall quality scores in all four REF panels (Figure 1). The associations were stronger for ChatGPT originality than for significance or rigour across the panels (Figure 2). The ChatGPT originality score difference was also larger than both other differences in 30 of the 34 UoAs, including every UoA in Panels A–C (Figure 3). Abstracts containing phrases from two or more novelty families received the highest originality scores in all four panels (Figure 5). These findings suggest that ChatGPT may be sensitive to novelty phrases rather than deducing novelty from the descriptions, although they do not show cause-and-effect: that the wording caused the higher scores rather than more novel research having been described with more novel language.

In answer to the second research question, abstracts containing novelty phrases were ranked more favourably by ChatGPT relative to the REF departmental proxy in 33 out of the 34 UoAs (Figure 4) and these differences were statistically significant in 26 UoAs. However, this does not show that ChatGPT is more sensitive to novelty wording than individual REF reviewers, because the REF scores used here were departmental averages.

In answer to the third research question, abstracts containing novelty phrases received higher mean ChatGPT overall quality scores in every REF panel, with the largest difference in Panel C (Figure 1). The originality-score difference was also larger than both the significance and rigour score differences in 30 of the 34 UoAs, including every UoA in Panels A–C (Figure 3). The four exceptions were UoAs 25, 29, 31 and 33 in Panel D, where only 11–57 abstracts contained an accepted novelty phrase. The pattern appeared across most subjects, although the results do not show directly that ChatGPT was significantly more affected in one panel than another. The results also do not show that adding such phrases would increase ChatGPT scores, although smaller experimental studies suggest that wording changes can affect LLM review scores (see introduction). Originality is a central REF quality criterion, so it is reasonable for ChatGPT to respond to evidence that research is original, especially because the models were prompted using the REF guidelines. The concern is whether explicit novelty claims are sometimes treated as evidence of research quality without the underlying contribution being assessed. Because only article titles and abstracts were provided to the models, LLMs such as ChatGPT may be partly guessing whether research is original or novel from this wording, and their scores may be influenced by simply emphasising that the research is new, original or novel. The same may be true if full text is entered since ChatGPT may naturally focus most on the concise summary of an abstract.

The results suggest but do not prove that stressing novelty in abstracts may provide an advantage when they are used for research evaluation. Although several small-scale experimental studies have suggested that reframing abstracts or strengthening novelty claims can increase LLM review scores (Kaneko, 2026; L. Li et al., 2026; M. Li et al., 2026; Yang et al., 2026), it is unclear whether such wording systematically influences LLM research-quality scores at a much larger scale across published journal articles. Hence, if LLMs are used to support research assessment practices such as the REF, their sensitivity to explicit claims and other linguistic features should be regularly assessed.

## References

Australian Research Council (2019). State of Australian University Research 2018–19: ERA national report. Introduction, "Definition of Research". https://dataportal.arc.gov.au/era/nationalreport/2018/pages/introduction/index.html?id=background

Howell, A., Wang, J., Du, L., Melkers, J., & Shah, V. (2025). Prestige over merit: An adapted audit of LLM bias in peer review. https://arxiv.org/abs/2509.15122

Huang, S., Huang, Y., Liu, Y., Luo, Z., & Lu, W. (2025). Are large language models qualified reviewers in originality evaluation? Information Processing & Management, 62(3), 103973.

Kaneko, M. (2026). Paraphrasing adversarial attack on LLM-as-a-reviewer. https://arxiv.org/abs/2601.06884

Kousha, K., & Thelwall, M. (2026a). Does ChatGPT score research quality differently by gender? https://arxiv.org/abs/2608.09552

Kousha, K., & Thelwall, M. (2026b). Which abstract stylistic features fool ChatGPT research evaluations? Scientometrics, 131(9), 6497–6518. https://doi.org/10.1007/s11192-026-05756-1

Langfeldt, L., Nedeva, M., Sörlin, S., & Thomas, D. A. (2020). Co-existing notions of research quality: A framework to study context-specific understandings of good research. *Minerva, 58*(1), 115–137.

Li, L., Zhang, Q., Davies, X., Qiu, J., & Gal, Y. (2026). Gaming AI-assisted peer reviews poses new risks to the scientific community. https://arxiv.org/abs/2606.10159

Li, M., Wang, C., Li, X., Zeng, X., Li, D., Shi, P., ... & Zhou, T. (2026). How can rhetoric reward-hack AI reviewers? Dissecting rhetorical sensitivity in AI-based peer review. https://arxiv.org/abs/2608.08975

Liu, N. F., Lin, K., Hewitt, J., Paranjape, A., Bevilacqua, M., Petroni, F., & Liang, P. (2024). Lost in the middle: How language models use long contexts. Transactions of the Association for Computational Linguistics, 12, 157-173.

Machado, D. (2026). Generative AI bias against scientific novelty: A cautionary tale from a small-sample evaluation of research proposals. Scientometrics, 131, 7479–7497. https://doi.org/10.1007/s11192-026-05792-x

Research Excellence Framework (REF). (2019). Panel criteria and working methods (REF 2019/02). https://2021.ref.ac.uk/media/1450/ref-2019_02-panel-criteria-and-working-methods.pdf

Thelwall, M. (2025a). Evaluating research quality with Large Language Models: An analysis of ChatGPT's effectiveness with different settings and inputs. Journal of Data and Information Science, 10(1), 7-25. https://doi.org/10.2478/jdis-2025-0011

Thelwall, M. (2025b). Is Google Gemini better than ChatGPT at evaluating research quality? Journal of Data and Information Science, 10(2), 1-5. https://doi.org/10.2478/jdis-2025-0014

Thelwall, M. (2026a). Do large language models know which published articles have been retracted? Library Management. https://doi.org/10.1108/LM-04-2026-0052

Thelwall, M. (2026b). Do large language models know basic facts about journal articles? Journal of Documentation, 82(2), 381–392. https://doi.org/10.1108/JD-11-2025-0330

Thelwall, M. (2026c). Do Large Language Models favour any research topics? https://arxiv.org/abs/2609.00323

Thelwall, M. (2026d). In which fields do ChatGPT scores align more closely with research quality than do citation rates? Journal of Data and Information Science. Advance online publication. https://www.degruyterbrill.com/document/doi/10.1515/jdis-2026-0058/html

Thelwall, M. (2026e). Will AI be overconfident about academic research findings when reliant on abstracts? https://arxiv.org/abs/2605.27392

Thelwall, M., & Ali, P. (2026). Is ChatGPT as reliable as individual reviewers assessing the quality of published journal articles from PDFs or titles and abstracts? *arXiv preprint arXiv:2607.25965*.

Thelwall, M., & Kurt, Z. (2025). Research evaluation with ChatGPT: Is it age, country, length, or field biased? Scientometrics, 130(10), 5323-5343. https://doi.org/10.1007/s11192-025-05393-0

Thelwall, M., & Mohammadi, E. (2026). Can small and reasoning large language models score journal articles for research quality and do averaging and few-shot help? Scientometrics, 1-33. https://doi.org/10.1007/s11192-026-05585-2

Thelwall, M., Kousha, K., Wilson, P., Makita, M., Abdoli, M., Stuart, E., ... & Cancellieri, M. (2023). Predicting article quality scores with machine learning: The UK Research Excellence Framework. Quantitative Science Studies, 4(2), 547-573.

Thelwall, M., Lehtisaari, M., Katsirea, I., Holmberg, K., & Zheng, E.-T. (2025). Does ChatGPT ignore article retractions and other reliability concerns? Learned Publishing, 38(4), e2018. https://doi.org/10.1002/leap.2018

Thelwall, M. & Yaghi, A. (2025) In which fields can ChatGPT detect journal article quality? An evaluation of REF2021 results. Trends in Information Management, 13 (1). pp. 1-29.

Universities of the Netherlands, Royal Netherlands Academy of Arts and Sciences, & Dutch Research Council. (2026). The strategy evaluation protocol 2027–2033. https://d2y7tnbi225lw.cloudfront.net/2026-03/Strategy-Evaluation-Protocol-%28SEP%29-2027-2033.pdf

Wu, W., Zhang, C., Bao, T., & Zhao, Y. (2025). SC4ANM: Identifying optimal section combinations for automated novelty prediction in academic papers. Expert Systems with Applications, 273, 126778. https://doi.org/10.1016/j.eswa.2025.126778

Yang, X., Sha, Z., Li, J., Yu, J., Sun, Y., Zhao, M., ... & Wang, Z. (2026). No hidden prompts needed! You can game AI peer review with presentation-only revisions. https://arxiv.org/abs/2606.13044

**Appendix A. LLM prompt for identifying novelty-claim phrases**

The following prompt was used with Gemma 3 12B to classify whether each three-word phrase was likely to indicate an explicit authorial originality or novelty claim. Each phrase was inserted after “PHRASE:”, and the model was instructed to return only 1 or 0.

*Decide whether this three-word phrase is a useful linguistic indicator that the sentence containing it very probably expresses an authorial novelty claim.*

*A novelty claim occurs when authors indicate that their research, method, finding, dataset, theory, approach, or contribution is new, original, previously unknown, unprecedented, or the first of its kind.*

*Classify the PHRASE itself, not merely whether you can imagine a sentence containing the phrase that also happens to contain a novelty claim elsewhere.*

*Return 1 only when the words in the phrase provide meaningful evidence of novelty-claim language.*

*Examples likely to be 1:*

*for the first*

*the first time*

*a novel approach*

*a new method*

*first study to*

*a previously unidentified*

*developed a new*

*first evidence of*

*provide a novel*

*Examples likely to be 0:*

*we introduce a*

*we propose a*

*results indicate that*

*in this study*

*we found that*

*has not previously*

*was associated with*

*the aim of*

*to our knowledge*

*not been reported*

*significantly higher than*

*data were collected*

*in patients with*

*distinct patterns of*

*Return ONLY:*

*1 = likely novelty indicator*

*0 = not a likely novelty indicator*

*PHRASE:*

**Supplementary Table S1. Accepted novelty phrases expressed as three-word phrases (3-grams) (N = 282). Phrases are listed alphabetically.**

| Accepted phrase | Accepted phrase | Accepted phrase |
|---|---|---|
| a first in | first demonstration of | open up new |
| a first step | first evidence of | original contribution to |
| a new algorithm | first evidence that | paper a novel |
| a new analytical | first example of | present a new |
| a new approach | first large scale | present a novel |
| a new concept | first report of | present the first |
| a new conceptual | first step towards | presents a new |
| a new design | first study to | presents a novel |
| a new family | first time a | presents the first |

| Accepted phrase | Accepted phrase | Accepted phrase |
|---|---|---|
| a new framework | first time and | previously unrecognized role |
| a new hybrid | first time in | propose a new |
| a new mechanism | first time that | propose a novel |
| a new method | first time the | proposes a new |
| a new methodology | first time this | proposes a novel |
| a new model | first time using | provide a new |
| a new paradigm | first time we | provide a novel |
| a new perspective | first to show | provide a unique |
| a new role | for a new | provide an alternative |
| a new route | for a novel | provide new insight |
| a new set | for the first | provide new insights |
| a new strategy | has not previously | provide novel insights |
| a new technique | here a new | provide the first |
| a new theoretical | here a novel | provides a new |
| a new therapeutic | identified a novel | provides a novel |
| a new type | identify a novel | provides a unique |
| a new way | in a first | provides new insights |
| a newly developed | in a new | provides the first |
| a novel algorithm | in the first | providing the first |
| a novel and | in this new | report a new |
| a novel approach | introduce a new | report a novel |
| a novel combination | introduce a novel | report the discovery |
| a novel framework | introduces a new | report the first |
| a novel hybrid | introduces a novel | represent a novel |
| a novel mechanism | is a new | represents a new |
| a novel method | is a novel | represents a novel |
| a novel methodology | is a unique | represents the first |
| a novel model | is superior to | results provide new |
| a novel multi | is the first | reveal a new |
| a novel regulator | makes an original | reveal a novel |
| a novel role | new approach for | shed new light |
| a novel target | new approach is | study a novel |
| a novel technique | new approach to | study provides new |
| a novel theoretical | new approaches to | that a new |
| a novel therapeutic | new data on | the first and |
| a novel way | new data set | the first application |
| a potential new | new evidence on | the first attempt |
| a previously uncharacterized | new form of | the first comprehensive |
| a previously undescribed | new forms of | the first demonstration |
| a previously unidentified | new insight into | the first detailed |
| a previously unknown | new insights into | the first direct |

| Accepted phrase | Accepted phrase | Accepted phrase |
|---|---|---|
| a previously unrecognized | new method for | the first empirical |
| a promising new | new method is | the first evidence |
| a unique data | new method of | the first example |
| a unique dataset | new method to | the first experimental |
| alternative approach to | new model of | the first in |
| an alternative method | new perspective on | the first large |
| an innovative approach | new role for | the first method |
| an original contribution | new set of | the first paper |
| an original dataset | new targets for | the first report |
| and a novel | new type of | the first results |
| and provide new | new types of | the first study |
| and the new | new understanding of | the first systematic |
| are the first | new way of | the first time |
| as a first | new ways of | the first to |
| as a novel | not been previously | the new approach |
| be a novel | not previously been | the new method |
| be the first | novel application of | the new model |
| by a new | novel approach for | the newly developed |
| by a novel | novel approach to | these novel findings |
| compared to previous | novel class of | this new approach |
| contrary to previous | novel combination of | this paper challenges |
| contrast to previous | novel framework for | this study advances |
| creation of new | novel insight into | to a novel |
| demonstrate a new | novel insights into | to develop new |
| demonstrate a novel | novel mechanism by | to identify novel |
| describe a new | novel mechanism for | to provide new |
| describe a novel | novel mechanism of | unique combination of |
| describes a new | novel method for | unique data set |
| describes a novel | novel method of | unique dataset of |
| design of new | novel method to | used a novel |
| develop a new | novel regulator of | using a new |
| develop a novel | novel target for | using a novel |
| developed a new | novel therapeutic strategy | using a unique |
| developed a novel | novel therapeutic target | using the new |
| development of new | novel therapeutic targets | was the first |
| development of novel | novelty of this | we create a |
| discovery of new | of a novel | we first propose |
| discovery of the | offer a new | we present new |
| drawing on original | offer new insights | we solve the |
| emergence of new | offers a new | with a new |
| findings provide new | offers an alternative | with a novel |

| **Accepted phrase** | **Accepted phrase** | **Accepted phrase** |
|---|---|---|
| findings provide novel | on a new | with the first |
| first application of | on a novel | with the new |
| first attempt to | on a unique | work a novel |